\documentclass[journal]{IEEEtran}
\usepackage{graphicx, soul}
\usepackage{soul}
\usepackage{amssymb}
\usepackage{slashbox}
\usepackage{cite}
\usepackage{subfigure}

\usepackage{amsmath}
\usepackage{xcolor}
\usepackage{footnote}

\newtheorem{Lemma}{Lemma}
\newtheorem{Theorem}{Theorem}

\begin{document}
%
\title{Performance Limits of Fluid Antenna Systems\thanks{The work is supported by EPSRC under grant EP/M016005/1.}}
%
%
%

\author{Kai-Kit Wong,~\IEEEmembership{Fellow,~IEEE,}
         Arman Shojaeifard,
        Kin-Fai Tong,~\IEEEmembership{Senior Member,~IEEE,}\\
        and~Yangyang Zhang

\thanks{K. Wong and K. Tong are with the Department of Electronic and Electrical Engineering, University College London, London WC1E 7JE, UK.}
\thanks{A. Shojaeifard is with BT Labs, Ipswich, UK.}
\thanks{Y. Zhang is with Kuang-Chi Institute of Advanced Technology, China.}
}

%
%

\markboth{Submitted to IEEE Communications Letters, 2020}%
{6G Wireless System}
%



\maketitle

\begin{abstract}
Fluid antenna represents a concept where a mechanically flexible antenna can switch its location freely within a given space. Recently, it has been reported that even with a tiny space, a single-antenna fluid antenna system (FAS) can outperform an $L$-antenna maximum ratio combining (MRC) system in terms of outage probability if the number of locations (or ports) the fluid antenna can be switched to, is large enough. This letter aims to study if extraordinary capacity can also be achieved by FAS with a small space. We do this by deriving the ergodic capacity, and a capacity lower bound. This letter also derives the level crossing rate (LCR) and average fade duration (AFD) for the FAS. 
\end{abstract}

\begin{IEEEkeywords}
Capacity, Fluid antennas, MIMO.
\end{IEEEkeywords}

%
\IEEEpeerreviewmaketitle

\vspace{-.05in}
\section{Introduction}
A main challenge in mobile phone design is the increasing limited physical dimensions. It is common practice that multiple antennas are only deployed, if they are sufficiently apart. The rule of thumb is that antennas should have a separation of at least $\frac{\lambda}{2}$ where $\lambda$ is the wavelength. However, this approach may need to be changed because the intuition that a tiny space does not have rich diversity is far from accurate. 

In \cite{Wong-2020}, a novel fluid antenna system (FAS) was investigated, where the antenna can be switched to one of $N$ fixed locations in a  linear space. The work was motivated by the emergence of mechanically flexible antennas, some of which are based on liquid metal antennas, e.g.,  \cite{Soh-2020,Hayes-2012,Shiroma-2013,Dey-2016} or ionized solutions \cite{Tong-2017,Tong-2018,Singh-2019}, while others utilize pixel antennas \cite{Murch-2014}. The beauty of `fluid' antennas is that an antenna is no longer fixed at a location but can be switched to a more favourable location if needed.

A key finding in \cite{Wong-2020} is that though space matters, a single-antenna FAS with a tiny space can achieve any arbitrarily small outage probability and surpass a multi-antenna maximum ratio combining (MRC) system, if $N$ is large enough. The aim of this letter is to examine if the promising outage performance of FAS can be translated into other benefits. In particular, this letter derives the level crossing rate (LCR), the average fade duration (AFD), and the ergodic capacity for the $N$-port FAS. One major contribution is a closed-form capacity lower bound for the FAS. Numerical results confirm the huge capacity gain unfolded from the diversity hidden in a small space of FAS.




\begin{figure}[ht!]
\centering
\includegraphics[width=.9\linewidth]{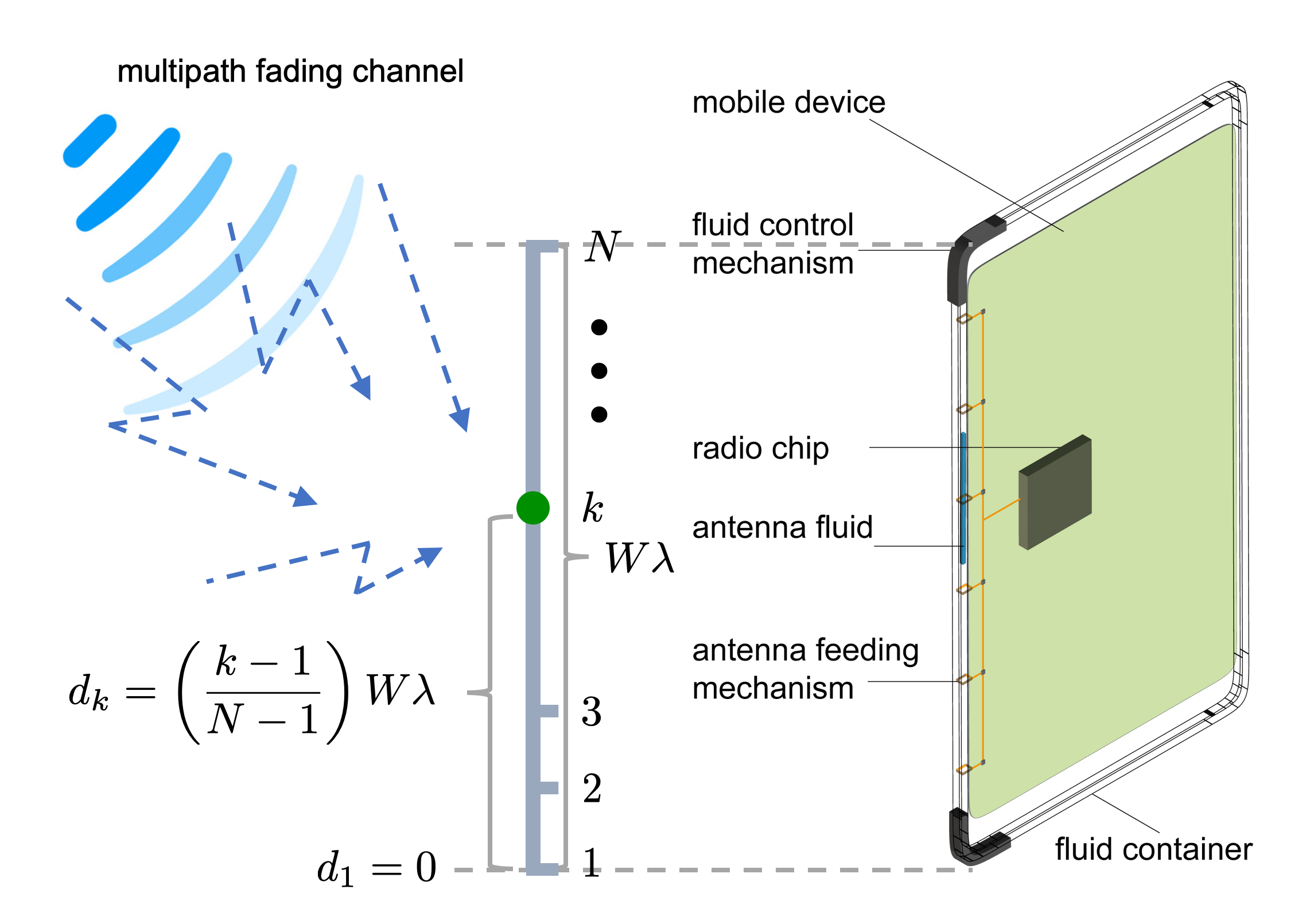}
\caption{The concept of FAS. In practice, the relocation of the antenna can be realized by mechanically moving liquid conductors using a microfluidic system, or electronically turning on or off radiating pixels.}\label{fig:fas}
\end{figure}

\section{FAS System Model}\label{sec_model}
We consider a FAS with linear space of $W\lambda$ where $\lambda$ is the wavelength, as shown in Figure \ref{fig:fas}. It is a single-antenna system where the antenna is mechanically flexible, using technologies such as microfluidic systems \cite{Dey-2016,Tong-2017,Tong-2018} or on-off pixels \cite{Murch-2014}, and can be switched to one of the $N$ preset locations along the space. We refer to the location as port, and the ports are evenly distributed over the space of $W\lambda$, sharing one RF chain.

The received signal at the $k$-th port can be written as
\begin{equation}
y_k=g_kx+\eta_k,
\end{equation}
in which $x$ is the transmitted symbol, $\eta_k$ is the additive white Gaussian noise (AWGN) with zero mean and variance of $\sigma_{\eta}^2$,  $g_k$ is the complex channel envelope, and $r_k=|g_k|$ is Rayleigh distributed, with the probability density function (pdf)
\begin{equation}\label{eqn:pdf-r}
p_{r_k}(r)=\frac{2r}{\sigma^2}e^{-\frac{r^2}{\sigma^2}},~\mbox{for }r_k\ge 0,
\end{equation}
with ${\rm E}[r_k^2]=\sigma^2$. Also, we define the average signal-to-noise ratio (SNR) as $\Gamma\triangleq\sigma^2\frac{{\rm E}[|x|^2]}{\sigma_\eta^2}=\sigma^2\Theta$.

As the antenna ports can be arbitrarily close to each other, they are spatially correlated. We parameterize the channels for the $N$-port FAS by
\begin{equation}\label{eqn:g-model}
\left\{\begin{aligned}
g_1&=\sigma x_0 +j\sigma y_0\\
&\vdots\\
g_k&=\sigma\left(\sqrt{1-\mu_k^2}x_2+\mu_k x_0\right)\\
&\quad\quad+j\sigma\left(\sqrt{1-\mu_k^2}y_2+\mu_k y_0\right),~\mbox{for }1<k\le N,
\end{aligned}\right.
\end{equation}
where $x_0,x_1,\dots,x_N,y_0,y_1,\dots,y_N$ are all independent Gaussian random variables with zero mean and variance of $\frac{1}{2}$, and $\{\mu_k\}$ are the parameters that can be chosen freely to control the correlation between the channels $\{g_k\}$. Assuming 2-D isotropic scattering and isotropic receiver ports on the FAS, the autocorrelation parameters are given by \cite{Wong-2020,Stuber-2002}
\begin{equation}\label{eqn:mu-condition}
\mu_k=J_0\left(\frac{2\pi(k-1)}{N-1}W\right),~\mbox{for }k=2,\dots,N,
\end{equation}
where $J_0(\cdot)$ is the zero-order Bessel function of the first kind.

In this letter, we assume that the FAS can always select the best port with the strongest signal for communication, i.e.,
\begin{equation}\label{eqn:rfas}
r_{\rm FAS}=\max\{r_1,r_2,\dots,r_N\}.
\end{equation}

%

\section{LCR and AFD}\label{sec_lcr}
LCR and AFD are two important characteristics to measure the rapidity and severity of fading. To analyze this, we need the time-derivatives of the signal envelopes, $\{\dot{r}_k\}_{\forall k}$. According to \cite{Jakes-1974}, for isotropic scattering, $\dot{r}_k$ is Gaussian distributed with zero mean and variance $\hat{\sigma}^2=2\pi^2\sigma^2f_m^2$ where $f_m=v/\lambda$ is the maximum Doppler shift for speed $v$. 

\begin{Theorem}
The LCR for an $N$-port FAS, for a level $r$, is given by
\begin{equation}\label{eqn:lcr}
N_{r_{\rm FAS}}(r)=\frac{2\sqrt{\pi}v}{\sigma\lambda}re^{-\frac{r^2}{\sigma^2}}.
\end{equation}
\end{Theorem}

\proof From \cite[(14)]{Takis-2002}, we have
\begin{align}
N_{r_{\rm FAS}}(r)&=\int_0^\infty \dot{r}p_{r_{\rm FAS}\dot{r}_{\rm FAS}}(r,\dot{r}) d\dot{r}\notag\\
&=\sum_{j=1}^N N_{r_{\rm FAS}|r_j}(r|r_j){\rm Prob}(r_j=\max\{r_k\})\notag\\
&=\frac{\hat{\sigma}p_r(r)}{\sqrt{2\pi}}\sum_{j=1}^N{\rm Prob}(r_j=\max\{r_k\}),\notag\\
&=\frac{\hat{\sigma}p_r(r)}{\sqrt{2\pi}},
\end{align}
where $p_r(r)$ is given by (\ref{eqn:pdf-r}). We then obtain (\ref{eqn:lcr}) after using (\ref{eqn:pdf-r}) and the fact that $\hat{\sigma}^2=2\pi^2\sigma^2f_m^2=2\pi^2\sigma^2v^2/\lambda^2$.
\endproof

\begin{Theorem}
The AFD for an $N$-port FAS, for a level $r$, is given by
\begin{equation}\label{eqn:afd}
\tau_{r_{\rm FAS}}(r)=\frac{{\rm Prob}(r_{\rm FAS}\le r)}{N_{r_{\rm FAS}}(r)},
\end{equation}
where the outage probability is given by \cite[Theorem 3]{Wong-2020} and $N_{r_{\rm FAS}}(r)$ is obtained by (\ref{eqn:lcr}).
\end{Theorem}

\proof The result comes from the definition of AFD.
\endproof


\section{Ergodic Capacity}\label{sec_rate}
In this section, we derive the ergodic capacity of the FAS in an integral form and obtain a lower bound in closed form. To do so, we find the following lemma useful.

\begin{Lemma}\label{lemma:ec}
The ergodic capacity can be obtained by
\begin{multline}\label{eqn:cap}
{\rm E}[\ln(1+\gamma)]=\int_0^\infty\left(\frac{1}{1+y}\right){\rm Prob}\left(r>\sqrt{\frac{y}{\Theta}}\right)dy,\\
({\rm in~nats/channel\text{-}use}),
\end{multline}
where $\gamma=r^2\Theta$ is the instantaneous SNR.
\end{Lemma}

\proof Using the fact that ${\rm E}[Y]=\int_0^\infty{\rm Prob}(Y>x)dx$, we get
\begin{align}
{\rm E}[\ln(1+\gamma)]&={\rm E}[\ln(1+r^2\Theta)]\notag\\
&=\int_0^\infty{\rm Prob}\left(\ln(1+r^2\Theta)>x\right)dx\notag\\
&=\int_0^\infty{\rm Prob}\left(r^2\Theta>e^x-1\right)dx.
\end{align}
Then by changing the variable $y=e^x-1$, we get (\ref{eqn:cap}).
\endproof

The following theorem establishes the exact ergodic capacity expression for the FAS in an analytical form.

\begin{Theorem}\label{theorem:ec}
The ergodic capacity of the FAS is given by
\begin{multline}\label{eqn:ec}
C_{\rm FAS}=\int_0^\infty\left(\frac{1}{1+y}\right)\left\{1-\int_0^\frac{y}{\Gamma}e^{-t}\times\right.\\
\left.\prod_{k=2}^N\left[1-Q_1\left(\sqrt{\frac{2\mu_k^2}{1-\mu_k^2}}\sqrt{t},\sqrt{\frac{2}{1-\mu_k^2}}\sqrt{\frac{y}{\Gamma}}\right)\right]dt\right\}dy,\\
({\rm in~nats/channel\text{-}use}),
\end{multline}
where $Q_1(\cdot,\cdot)$ denotes the first-order Marcum $Q$-function.
\end{Theorem}

\proof Using Lemma \ref{lemma:ec} and noting that
\begin{equation}
{\rm Prob}\left(r>\sqrt{\frac{y}{\Theta}}\right)=1-{\rm Prob}\left(r_{\rm FAS}\le\sqrt{\frac{y}{\Theta}}\right),
\end{equation}
we obtain (\ref{eqn:ec}) after substituting $\Gamma=\sigma^2\Theta$.
\endproof

Theorem \ref{theorem:ec} provides the exact result for the ergodic capacity but it is difficult to gain any insight. We now try to establish a lower bound in a closed form so that the capacity benefit for the spatially correlated ports can be quantified.

\begin{Lemma}\label{lemma:Q1-lb}
For $0<\alpha<\beta$ and large $\beta$, we have the following lower bound for $Q_1(\alpha,\beta)$:
\begin{equation}\label{eqn:Q1-lb}
Q_1(\alpha,\beta)\gtrsim\varrho\sqrt{\frac{\beta}{\alpha}}e^{-\frac{\kappa}{2}(\beta-\alpha)^2},
\end{equation}
where $\kappa$ is any positive constant greater than one, and $\varrho\triangleq\frac{e^\frac{1}{\pi(\kappa-1)+2}}{2\kappa}\sqrt{\frac{(\kappa-1)(\pi(\kappa-1)+2)}{\pi}}$. Note that $0<\varrho<0.5$.
\end{Lemma}

\proof See \cite[Lemma 6]{Wong-2020}.
\endproof

\begin{Theorem}\label{theorem:qn-lb}
Defining the service probability of an $N$-port FAS, for a level $y$, as
\begin{equation}
q_N(y)\triangleq{\rm Prob}\left(\left.r_{\rm FAS}>\sqrt{\frac{y}{\Theta}}\right|N\right),
\end{equation}
it can be lower-bounded by
\begin{equation}\label{eqn:qn}
q_N(y)\ge\sum_{\ell=1\atop {1\le k_1,\dots,k_\ell\le N\atop k_{\ell_1}\ne k_{\ell_2}}}^N(-1)^{\ell+1}\varepsilon_{k_1}\varepsilon_{k_2}\cdots\varepsilon_{k_\ell}
\end{equation}
where
\begin{equation}\label{eqn:ek}
\varepsilon_k\triangleq\left\{\begin{aligned}
\frac{\varrho}{\sqrt{|\mu_k|}}e^{-\frac{\kappa}{1-\mu_k^2}\left(\frac{y}{\Gamma}\right)}, &~\mbox{if }k>1~\mbox{and }|\mu_k|>\varrho^2\\
e^{-\frac{\kappa}{1-\mu_k^2}\left(\frac{y}{\Gamma}\right)}, &~\mbox{if }k>1~\mbox{and }|\mu_k|\le \varrho^2\\
e^{-\frac{y}{\Gamma}}, &~\mbox{if }k=1,
\end{aligned}\right.
\end{equation}
in which $\kappa>1$ and $0<\varrho<0.5$ are defined in Lemma \ref{lemma:Q1-lb}.
\end{Theorem}

\proof First of all, define the outage probability so that
\begin{equation}
p_N(y)={\rm Prob}\left(\left.r_{\rm FAS}\le\sqrt{\frac{y}{\Theta}}\right|N\right)=1-q_N(y).
\end{equation}
A lower bound of $q_N(y)$ can be obtained by developing an upper bound for $p_N(y)$. Denoting $\alpha_k=\sqrt{\frac{2\mu_k^2}{1-\mu_k^2}}$ and $\beta_k=\sqrt{\frac{2}{1-\mu_k^2}}\sqrt{\frac{1}{\Gamma}}$, we can find that
\begin{align}
p_N(y)
&=p_{N-1}(y)-\int_0^\frac{y}{\Gamma}Q_1\left(\alpha_N\sqrt{t},\beta_N\sqrt{y}\right)\times\notag\\
&\quad\quad\quad\quad e^{-t}\prod_{k=2}^{N-1}\left[1-Q_1\left(\alpha_k\sqrt{t},\beta_k\sqrt{y}\right)\right]dt.\label{eqn:qn-2}
\end{align}
As $\alpha_N\sqrt{t}\le\beta_N\sqrt{y}$, we use Lemma \ref{lemma:Q1-lb} to get
\begin{equation}
Q_1(\alpha_N\sqrt{t},\beta_N\sqrt{y})\gtrsim\frac{\varrho}{\sqrt{|\mu_N|}}\frac{\left(\frac{y}{\Gamma}\right)^{0.25}}{t^{0.25}}e^{-\frac{\kappa}{1-\mu_N^2}\left(\sqrt{\frac{y}{\Gamma}}-\mu_N\sqrt{t}\right)^2}.
\end{equation}
As $0\le t\le\frac{y}{\Gamma}$, we can further have the following lower bound
\begin{equation}
Q_1(\alpha_N\sqrt{t},\beta_N\sqrt{y})\gtrsim\frac{\varrho}{\sqrt{|\mu_N|}}e^{-\frac{\kappa}{1-\mu_N^2}\left(\frac{y}{\Gamma}\right)}\equiv\varepsilon_N.
\end{equation}
Now, using this result inside the integration of (\ref{eqn:qn-2}), we have
\begin{align}
p_N(y)&\le (1-\varepsilon_N)p_{N-1}(y)\notag\\
&\le (1-\varepsilon_N)(1-\varepsilon_{N-1})p_{N-2}(y)\notag\\
&\vdots\notag\\
&\le p_1(y)\prod_{k=2}^N(1-\varepsilon_k).\label{eqn:pn-ub2}
\end{align}
As a result, we can obtain a lower bound for $q_N(y)$ by
\begin{equation}\label{eqn:qn-lb}
q_N(y)\ge 1-p_1(y)\prod_{k=2}^N(1-\varepsilon_k).
\end{equation}
Note that $p_1(y)=1-e^{-\frac{y}{\Gamma}}=1-\varepsilon_1$. If we then expand the product in (\ref{eqn:qn-lb}), then we have
\begin{equation}
\prod_{k=1}^N(1-\varepsilon_k)=1+\sum_{\ell=1\atop {1\le k_1,\dots,k_\ell\le N\atop k_{\ell_1}\ne k_{\ell_2}}}^N(-1)^{\ell}\varepsilon_{k_1}\varepsilon_{k_2}\cdots\varepsilon_{k_\ell}.
\end{equation}
Using the above result on (\ref{eqn:qn-lb}) gives the desired result. Note that (\ref{eqn:pn-ub2}) is valid only if $\frac{\varrho}{\sqrt{|\mu_k|}}e^{-\frac{\kappa}{1-\mu_k^2}\left(\frac{y}{\Gamma}\right)}\le 1$. Therefore, the definition (\ref{eqn:ek}) is introduced to guarantee that.
\endproof


\begin{Theorem}\label{theorem:ec-lb}
The ergodic capacity for an $N$-port FAS in (\ref{eqn:ec}) is lower bounded by
\begin{multline}\label{eqn:ec-lb}
C_{\rm FAS}\ge C_{\rm LB}
=\sum_{\ell=1\atop {1\le k_1,\dots,k_\ell\le N\atop k_{\ell_1}\ne k_{\ell_2}}}^N(-1)^{\ell+1}a_{k_1}\cdots a_{k_\ell}\times\\
 e^{\sum_{j=1}^\ell b_{k_j}}\Gamma_{\rm inc}\left(0,\sum_{j=1}^\ell b_{k_j}\right),
\end{multline}
where $\Gamma_{\rm inc}(x,y)$ is the upper incomplete Gamma function,
\begin{equation}
a_k\triangleq\left\{\begin{aligned}
1 &~\mbox{if }k=1~\mbox{or }|\mu_k|\le\varrho^2,\\
\frac{\varrho}{\sqrt{|\mu_k|}} &~\mbox{if }k\ne 1~\mbox{and }|\mu_k|>\varrho^2,
\end{aligned}\right.
\end{equation}
and
\begin{equation}
b_k\triangleq\left\{\begin{aligned}
\frac{1}{\Gamma} &~\mbox{if }k=1,\\
\frac{\kappa}{1-\mu_k^2}\left(\frac{1}{\Gamma}\right) &~\mbox{if }k\ne 1.
\end{aligned}\right.
\end{equation}
\end{Theorem}

\proof From Lemma \ref{lemma:ec}, we have
\begin{equation}
C_{\rm FAS}=\int_0^\infty\frac{q_N(y)}{1+y}dy.
\end{equation}
Using the lower bound for $q_N(y)$ in Theorem \ref{theorem:qn-lb}, we get
\begin{align}
C_{\rm FAS}&\ge \sum_{\ell=1\atop {1\le k_1,\dots,k_\ell\le N\atop k_{\ell_1}\ne k_{\ell_2}}}^N(-1)^{\ell+1}\int_0^\infty\frac{\varepsilon_{k_1}\varepsilon_{k_2}\cdots\varepsilon_{k_\ell}}{1+y}dy\notag\\
&=\sum_{\ell=1\atop {1\le k_1,\dots,k_\ell\le N\atop k_{\ell_1}\ne k_{\ell_2}}}^N(-1)^{\ell+1}a_{k_1}\cdots a_{k_\ell}\times\notag\\
&\quad\quad\quad\quad\quad\quad\int_0^\infty\frac{e^{-b_{k_1}y}\cdots e^{-b_{k_\ell}y}}{1+y}dy.
\end{align}
Using the fact that $\int_0^\infty\frac{e^{-cy}}{1+y}dy=e^c\Gamma_{\rm inc}(0,c)$ gives the final result (\ref{eqn:ec-lb}), which completes the proof.
\endproof

\section{Numerical Results}\label{sec_result}
\begin{figure}[]
\centering
\includegraphics[width=\linewidth]{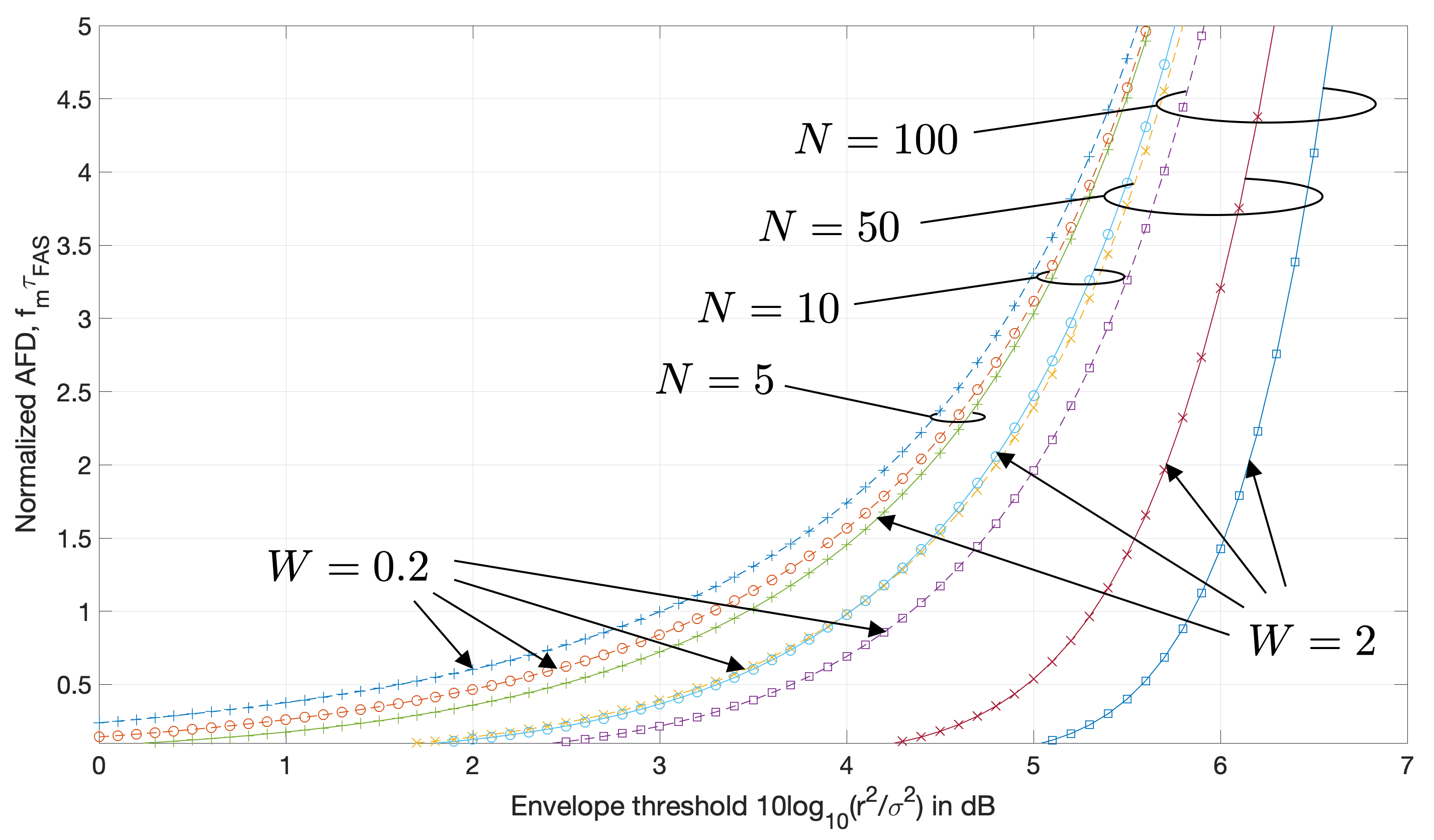}
\caption{The normalized AFD results against the level for various $W$ and $N$.}\label{fig:afd}
\end{figure}

In this section, we provide the numerical results for the AFD and ergodic capacity under different settings. Figure \ref{fig:afd} shows the AFD results against the signal envelope level for the FAS with different $N$ and size $W$. As expected, if the level is large, AFD eventually grows rapidly without bound. What is crucial is when this starts to happen. The results reveal that $N$ helps increase the signal level with practically zero AFD. Also,  the size $W$ has a positive impact on the AFD. If $W$ is larger, more diversity potentially exists and AFD can be reduced.

\begin{figure}[]
\begin{center}
\subfigure[Ergodic capacity versus number of ports when $\Gamma=10$dB]{\includegraphics[width=.88\linewidth]{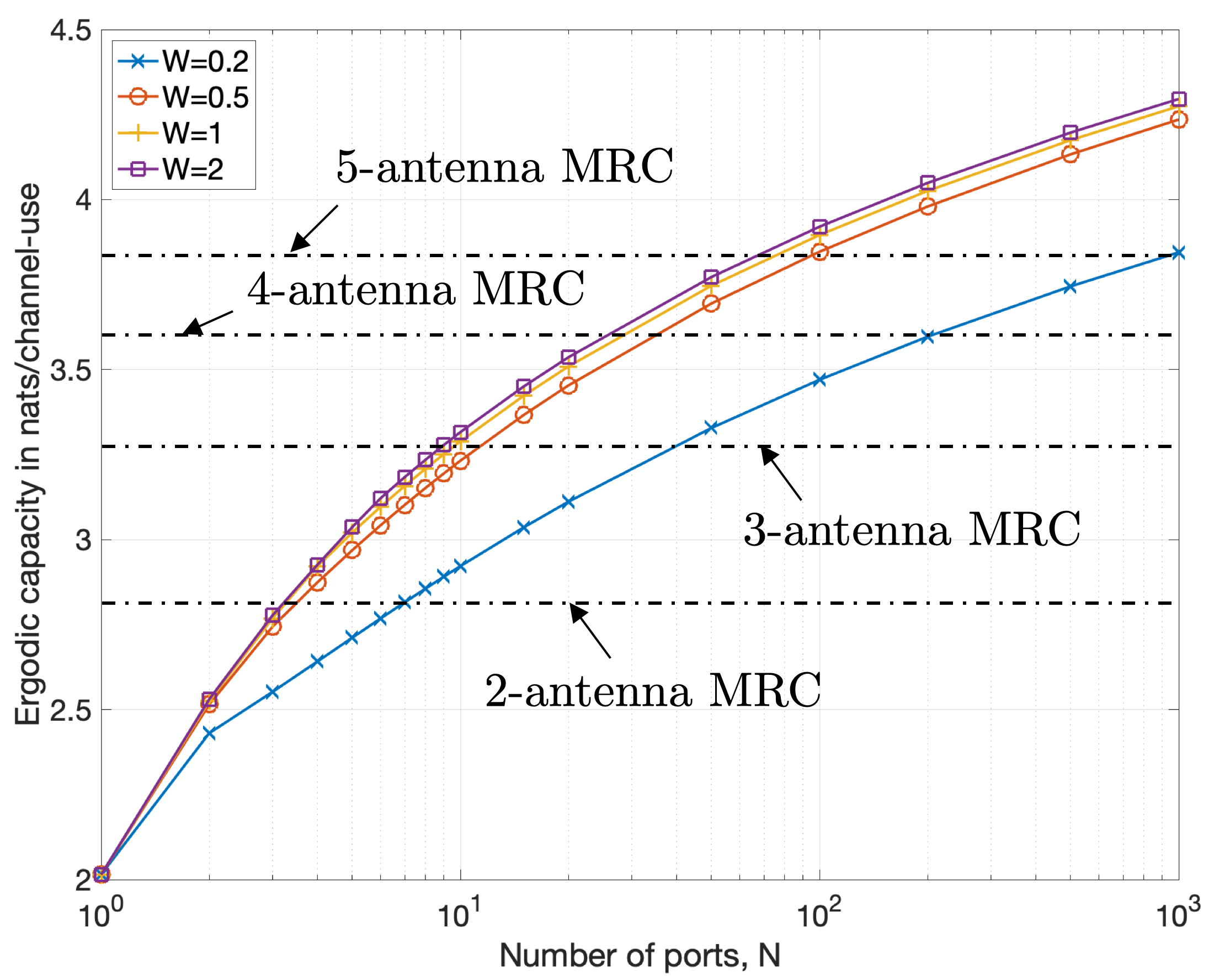}\label{fig:ecVSn}}
\subfigure[Ergodic capacity versus size when $\Gamma=10$dB]{\includegraphics[width=\linewidth]{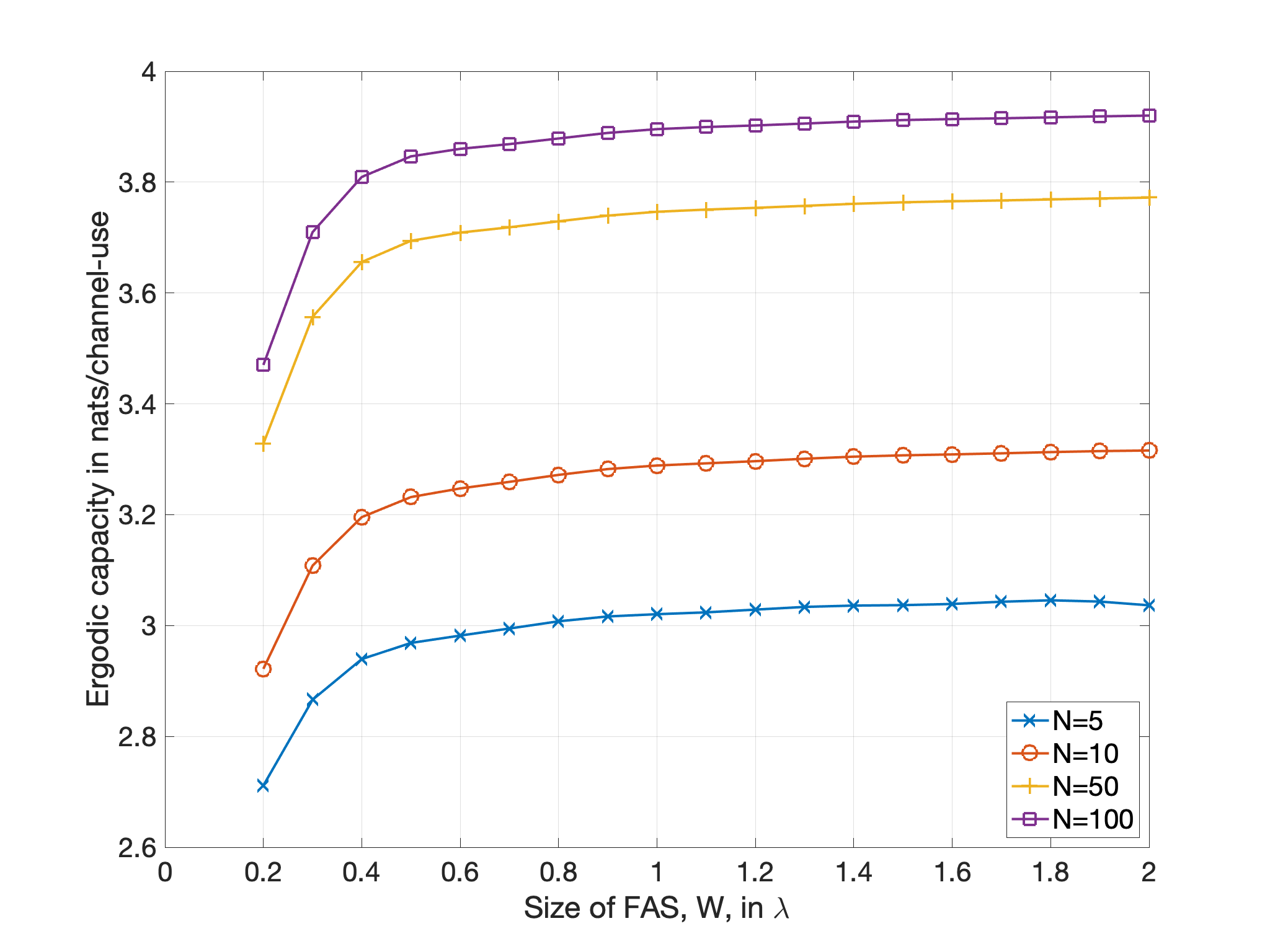}\label{fig:ecVSw}}
\subfigure[Ergodic capacity versus the lower bound with $W=1$]{\includegraphics[width=.88\linewidth]{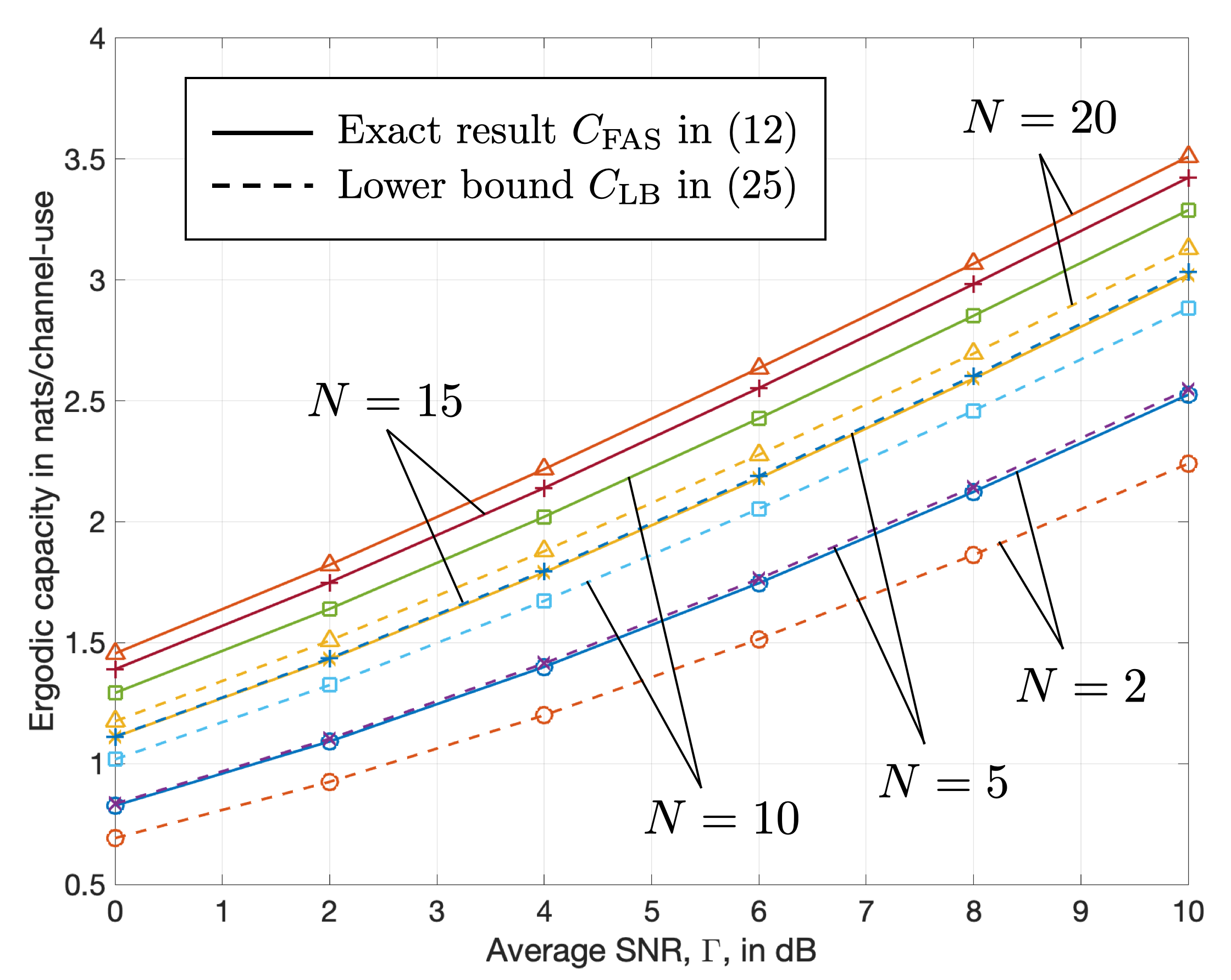}\label{fig:ecVSsnr}}
\caption{Ergodic capacity results for the FAS.}\label{fig:ec}
\end{center}
\end{figure}

Results in Figure \ref{fig:ec} are provided for the ergodic capacity of the FAS. Figure \ref{fig:ecVSn} illustrates how the capacity of the FAS scales with $N$. As we can see, capacity continues to increase with $N$ even when $W$ is very small. Also, $W=0.5$ appears to have the biggest jump in performance, as further increase in $W$ only contributes little. This is further confirmed by the results in Figure \ref{fig:ecVSw}, where it is shown that capacity plateaus when $W$ reaches $1$. This reveals that $N$ is a more important factor than $W$ in the FAS. Remarkably, results also demonstrate that the capacity of a single-antenna FAS can obtain that of a multi-antenna MRC system with independent fading. In particular, $N=10$ will be enough for the FAS with $W\ge 0.5$ to approach the capacity for a 3-antenna MRC system. More capacity is also possible if $N$ continues to increase.

Lastly, results in Figure \ref{fig:ecVSsnr} evaluate the lower bound in Theorem \ref{theorem:ec-lb} against the exact capacity for different values of SNR. The results indicate that although the lower bound is not particularly tight, it accurately picks up how the capacity grows with the SNR. Also, the lower bound scales very well with $N$. As we can see, the gap between the bound and the exact result remains similar if $N$ increases.

\section{Conclusion}\label{sec_con}
Following the emergence of mechanically flexible antennas, this letter studied if the diversity of FAS with a small space can be translated into substantial capacity gain. To this end, we first derived the exact ergodic capacity in an analytical form, and then established a capacity lower bound that was able to reveal how the capacity scales with the system parameters. We concluded that a single-antenna FAS, with a small space, can achieve the capacity that a multi-antenna MRC system has.

\ifCLASSOPTIONcaptionsoff
  \newpage
\fi



\begin{thebibliography}{1}

\bibitem{Wong-2020}
K. K. Wong, A. Shojaeifard, K. F. Tong, and Y. Zhang, ``Fluid antenna systems,'' [Online] arXiv:2005.11561 [cs.IT].

\bibitem{Soh-2020}
K. N. Paracha, A. D. Butt, A. S. Alghamdi, S. A. Babale, and P. J. Soh, ``Liquid metal antennas: Materials, fabrication and applications,'' {\em Sensors 2020}, 20, 177.
\bibitem{Hayes-2012}
G. J. Hayes, J.-H. So, A. Qusba, M. D. Dickey, and G. Lazzi, ``Flexible liquid metal alloy (EGaIn) microstrip patch antenna,'' {\em IEEE Trans. Antennas Propag.}, vol. 60, no. 5, pp. 2151--2156, May 2012.
\bibitem{Shiroma-2013}
A. M. Morishita, C. K. Y. Kitamura, A. T. Ohta, and W. A. Shiroma, ``A liquid-metal monopole array with tunable frequency, gain, and beam steering,'' {\em IEEE Antennas Wireless Propag. Lett.}, vol. 12, pp. 1388--1391, 2013.
\bibitem{Dey-2016}
A. Dey, R. Guldiken, and G. Mumcu, ``Microfluidically reconfigured wideband frequency-tunable liquid-metal monopole antenna,'' {\em IEEE Trans. Antennas Propag.}, vol. 64, no. 6, pp. 2572--2576, Jun. 2016.

\bibitem{Tong-2017}
C. Borda-Fortuny, K. F. Tong, A. Al-Armaghany, and K. K. Wong, ``A low-cost fluid switch for frequency-reconfigurable Vivaldi antenna,'' {\em IEEE Antennas Wireless Prop. Lett.}, vol. 16, pp. 3151--3154, Nov. 2017.
\bibitem{Tong-2018}
C. Borda-Fortuny, K. F. Tong, and K. Chetty, ``Low-cost mechanism to reconfigure the operating frequency band of a Vivaldi antenna for cognitive radio and spectrum monitoring applications,'' {\em IET Microwaves, Antennas \& Propag.}, vol. 12, no. 5, pp. 779--782, 2018.
\bibitem{Singh-2019}
A. Singh, I. Goode, and C. E. Saavedra, ``A multistate frequency reconfigurable monopole antenna using fluidic channels,'' {\em IEEE Antennas Wireless Propag. Lett.}, vol. 18, no. 5, pp. 856--860, May 2019.

\bibitem{Murch-2014}
S. Song, and R. D. Murch, ``An efficient approach for optimizing frequency reconfigurable pixel antennas using genetic algorithms," {\em IEEE Trans. Antennas Propag.}, vol. 62, no. 2, pp. 609--620, Feb. 2014.

\bibitem{Stuber-2002}
G. L. St$\ddot{\rm u}$ber, {\em Principles of Mobile Communication}, Second Edition, Kluwer Academic Publishers, 2002.

\bibitem{Jakes-1974}
W. C. Jakes, {\em Microwave Mobile Communications}. New York: Wiley, 1974.
\bibitem{Takis-2002}
C.-D. Iskander, and P. T. Mathiopoulos, ``Analytical level crossing rates and average fade durations for diversity techniques in Nakagami fading channels,'' {\em IEEE Trans. Commun.}, vol. 50, no. 8, pp. 1301--1309, Aug. 2002.

\end{thebibliography}
\end{document}